\documentclass[a4paper,12pt]{article}
\usepackage{amsmath,amsfonts,amssymb}
\newcommand*{\raD}{\stackrel{\text{D}}{\rightarrow}}

\begin{document}

\begin{center}
{\Large\bf Local duality symmetry in gauge field theories}
\end{center}

\begin{center}
{\large T. Z. Seidov}
\end{center}

\begin{center}
{\it Saint Petersburg State University, 7/9 Universitetskaya nab.,
St.Petersburg, 199034, Russia}
\end{center}
\begin{abstract}
    We develop the idea of local duality symmetry (LDS) in gauge field theories. Using Clifford algebra techniques we construct dually invariant scalar Lagrangian of electrodynamics in the presence of sources and demonstrate that in tensor formalism it is exactly the same as the usual one. Then we localize the duality symmetry with two possible options for the appearing pseudovector field - massive and massless. The first option might be interpreted as a candidate for dark matter. Perspectives of the application of LDS in QCD are briefly discussed. 
\end{abstract}
\begin{section}{Introduction}
\subsection{Duality transform}
	It has been known for a long time that free field Maxwell's equations $\partial_\mu F^{\mu\nu}=0,  \partial_\mu \star {F}^{\mu\nu}=0,$ where, as usual, $F_{\mu\nu}=\partial_\mu a_\nu - \partial_\nu a_\mu$ is electromagnetic strength tensor ($\star F_{\mu\nu}=1/2\epsilon_{\mu\nu \alpha \beta }F^{\alpha \beta }$), $a_\mu$ is 4-potential, are invariant under the so-called duality transforms \cite{Shnir2005MagneticMonopoles}
    \begin{eqnarray}
    F^{\mu\nu}\longrightarrow F^{\mu\nu} \cos(\theta) + \star F^{\mu\nu} \sin(\theta), \nonumber \\
    \star F^{\mu\nu}\longrightarrow \star F^{\mu\nu} \cos(\theta) - F^{\mu\nu} \sin(\theta). \label{ffdt}
    \end{eqnarray}
   The Lagrangian for such field equations can be written in form of $\mathcal{L}=F_{\mu\nu}^2$. After transform (\ref{ffdt}) it will be $\mathcal{L}=\cos(2\theta)F_{\mu\nu}^2+\sin(2\theta)\star{F}_{\mu\nu}F^{\mu\nu}$, but since the last term is the full divergence and multiplication by a constant does not change variation of action, duality transform for a free field might be considered as a Noether's symmetry, because variation of action still equals zero after transform.
    However, in the presence of charged particles, to maintain the symmetry of equations a magnetic charge is needed. Then the transform would be completed with transformations of current densities (magnetic and electric), or equivalently, of charges:
    \begin{eqnarray}
  e\longrightarrow e\cos(\theta) + g\sin(\theta), \nonumber \\
  g\longrightarrow g \cos(\theta) - e\sin(\theta). \label{cdt}
    \end{eqnarray}The standard electrodynamics Lagrangian $\mathcal{L}=-\frac{1}{4}F_{\mu\nu}^2-e j_\mu a^\mu$ ($e$ is the electric charge and $j_\mu$ is the current density) is not symmetrical under duality transform even in sense of Noether's theorem. As it is shown in the analysis done by Katz in \cite{Katz1965}, experimentally it is impossible to  prove whether all charged particles have only electric charge or both electric and magnetic charges (dyons) with universal ratio $g/e=\tan(\theta)$, which is fundamentally due to the fact that the stress-energy tensor of electromagnetic field is invariant under duality transform, and only fixation of $\theta=0$ leads to the well-known Maxwell's electrodynamics. In view of this statement a question appears whether it is possible to write dually symmetrical Lagrangian and is it possible to localize the dual symmetry (most commonly with the appearance of axion-like pseudoscalar field). A brief history of such attempts can be found in \cite{Tiwari2015AxionPerspective} and in the recent article by Visinelli \cite{Visinelli2011DualElectrodynamics}, as well as his own approach.
    \par However, the Lagrangian proposed in \cite{Visinelli2011DualElectrodynamics} as a dual invariant for a free field contains $F_{\mu\nu}^2+(\star F_{\mu\nu})^2$, which is zero in the situation described by Katz (dyon case). Moreover, he puts axion terms into Lagrangian manually, not by localizing the symmetry. Tiwari \cite{Tiwari2015AxionPerspective} localized duality symmetry in Sudbery's dually invariant 4-vector Lagrangian formalism described in \cite{Sudbery1986AField}, generalizing it for the case of present sources. We will concentrate on the dyon case because taking into account \cite{Katz1965}, it seems natural to demand dual symmetry. Also, we will do it in the scalar Lagrangian formalism. 
\subsection{Clifford-Dirac algebra}
The Clifford-Dirac algebra $Cl_\mathbb{C}(1,3)$, also known as the gamma-matrices algebra, has proven itself as a indispensable tool in physics. The defining equation for its generators is $\gamma^\nu \gamma^\mu + \gamma^\mu \gamma^\nu=2g^{\mu \nu}$. We mention briefly some properties and difficulties that will be used below. There is a special matrix $\gamma_5=i\gamma^0 \gamma^1 \gamma^2 \gamma^3$ that anti-commutes with each of $\gamma_\mu$. Hermitian conjugation acts in the following way: $(\gamma^\mu)^\dagger=g^{0\mu}\gamma^\mu$ (no summation), $\gamma_5^\dagger=\gamma_5$.  For any algebra element $A=\gamma_\mu\gamma_\nu ... \gamma_\alpha$ inversion operation is defined as inversion of multiplication order: $\widetilde A=\gamma_\alpha...\gamma_\nu \gamma_\mu$. For any complex (or real) number $\alpha$ we will use the notations: $\alpha$ - (Clifford) scalar, $\alpha\gamma_\mu$ - (Clifford) vector, $\alpha\gamma_\mu\gamma_\nu$ -(Clifford) bivector, $\alpha\gamma_\mu\gamma_\nu\gamma_\alpha$ -(Clifford) pseudovector, $\alpha \gamma_5$ - (Clifford) pseudoscalar. For any pseudoscalar we can define $*$ operation analogous to complex conjugation: $(\alpha \gamma_5)^*=-\alpha \gamma_5$. The usual trace identities fo gamma-matrices will be exploited. 
The Clifford algebra allows for an elegant formulation of electrodynamics, a short review is given in ref. \cite{Hestenes2015Space-TimeAlgebra}. In particular, with the use of Clifford algebra, Maxwell's equations can be written in a form of one equation \cite{Hestenes2015Space-TimeAlgebra}
\begin{eqnarray}
\partial F=ej , \label{max}
\end{eqnarray}here and henceforth, $j=j_\mu \gamma^\mu$, $a=a_\mu \gamma^\mu$, $\partial=\partial_\mu \gamma^\mu$, $F=\partial a - \widetilde{\partial a}=F_{\mu\nu}\gamma^\mu \gamma^\nu$. It can be split into common Maxwell's equation by making linear combinations of this equation $\gamma_5$( \ref{max} )$\gamma_5 \pm$ (\ref{max}). There exists \cite{Waldyr2016TheEquations} an expanded equation for the case of present magnetic 4-current represented as a pseudovector $k$: $\partial F=ej+gk$ . Some resources, including \cite{Waldyr2016TheEquations}, even mention the possibility of duality transform as a pseudoscalar phase rotation. 
As it will be shown, Clifford algebra provides excellent formalism for duality transform and it's localization.
\end{section} 
\begin{section}{General idea}
	The duality transforms (\ref{ffdt}) and (\ref{cdt}) may be rewritten using $-i\gamma_5 F=\star F$ as
 \begin{eqnarray}
 F\longrightarrow e^{-i\gamma_5 \theta}F, \nonumber \\
-i\gamma_5 F=\star F\longrightarrow e^{-i\gamma_5\theta}\star F, \nonumber \\
e \longrightarrow e^{-i\gamma_5\theta}e, \nonumber \\
-i\gamma_5 g \longrightarrow e^{-i\gamma_5\theta}(- i\gamma_5g).
 \end{eqnarray}Our first idea is to construct dually and gauge invariant, Lorentz-scalar Lagrangian that leads to Maxwell's equations, introducing parameter $q=e'-i\gamma_5g$, with a constraint $q=e \cdot \exp({i\gamma_5 \alpha)}$ (dyon condition) as a generalized electromagnetic charge, and generalized 4-potential $A=A_e-i\gamma_5A_g$, with a constraint $A=e^{i\gamma_5\beta} a$, or, equivalently (it can be obtained by squaring both parts of the constraint) $A_g^2+A_e^2=a^2$ and $A_{g\mu} A_{e\nu} -A_{g\nu}A_{e\mu}=0 \Rightarrow A_{g\mu} \propto A_{e\nu}$  , which are tensor constraints that prevent growth of degrees of freedom while using two-potential electrodynamics for every fixed $\beta$. Why should we choose charge and 4-potential? Let us start with the interaction term  $ej_\mu a^\mu$. If a charge or a current is dually transformed, we need to transform 4-potential in order to compensate such transformation. Thus, out of strength tensor and 4-potential we must choose the last one to transform. Then we must think how to compensate transform of $F^2$. We can either put dual invariant of $F$ that does not change dynamics instead of it or multiply $F^2$ by some dimensionless unimodular function $U(f_1(j, \theta), A, f_2(e, \theta))$, that is erased should we set $\theta=0$. There is no such invariant, because a minimal dual non-zero scalar invariant in our case is quadratic in $F^2$: $I=\frac{1}{4}[(F_{\mu\nu}^2)^2+(F_{\mu\nu}\star F^{\mu\nu})^2]$. And $U$ cannot depend on $j$ and $A$, because otherwise it would influence variation procedure as they depend on coordinates. Thus we choose charge.
\end{section}

\begin{section}{Dually symmetrical electrodynamics}
\par Consider the usual Lagrangian of electrodynamics without Dirac term in form of 
\begin{eqnarray}
\mathcal{L}=-\frac {1}{4} F_{\mu\nu}F^{\mu\nu}- e j^{\mu} a_{\mu}. \label{L0}
\end{eqnarray}Using trace operation one may rewrite it using Clifford-Dirac algebra elements and operations
\begin{eqnarray}
\mathcal{L}=\frac{1}{4} tr \left[ - ej a + \frac{1}{8}  F^2 \right] \label{L01}
\end{eqnarray}
	Now, if one considers all of these variables as elements of Clifford-Dirac algebra one may expand the range of values for charge and 4-potential giving them a pseudoscalar phase. Our new charge $q=e \cdot \exp{(i \gamma_5\alpha)}$ and 4-potential $A=\exp{(i \gamma_5\beta)} a$ are thus sums of scalar and pseudoscalar, and vector and pseudovector respectively. Physically, it means that all particles are now dyons, and as we will see from the equation (\ref{eq}), pseudoscalar part of the charge plays the role of magnetic charge. One must be careful now because charge is no longer commutative with any of gamma-matrices. Thus, we must define its position in our Lagrangian manually. However, its position doesn't change any further conclusions as a change of position in definition of Lagrangian always can be compensated by redefinition of duality transform. We may also multiply the free field term by $\frac{q^* q}{q^2}=\frac{q*}{q}$ which will provide us with a new symmetry, allowing to compensate phase of $\mathcal{F}^2$, while also vanishing if we get back to initial variables. Thus, the new Lagrangian is
\begin{eqnarray}
\mathcal{L}=\frac{1}{4} tr \left[ - jqA + \frac{1}{8} \frac{q^*}{q}  \mathcal{F}^2 \right]. \label{L}
\end{eqnarray}
$\mathcal{F}$ is defined as $ \mathcal{F}=\partial A - \widetilde{\partial A}$. After taking trace, operation it can be written in old variables as
\begin{eqnarray}
\mathcal{L}=\frac{1}{4} tr \left[ - ej a\exp{(-i\gamma_5(\alpha+\beta))} +  \frac{1}{8} F^2 \exp{(-i2 \gamma_5(\alpha+\beta))}\right]= \nonumber \\ =- \cos(\alpha+\beta) e j^{\mu} a_{\mu}-\cos(2(\alpha+\beta))\frac {1}{4} F_{\mu\nu}F^{\mu\nu} - \nonumber \\ - \sin(2(\alpha+\beta))\epsilon_{\mu\nu\rho\sigma}F^{\mu\nu}F^{\rho\sigma}.
\end{eqnarray}
Or, introducing the notation $\alpha+\beta=\xi$,
\begin{eqnarray}
\mathcal{L}=-\cos(\xi) e j^{\mu} a_{\mu}-\cos(2\xi)\frac {1}{4} F_{\mu\nu}F^{\mu\nu} - \sin(2\xi)\epsilon_{\mu\nu\rho\sigma}F^{\mu\nu}F^{\rho\sigma}.
\end{eqnarray}
From this form we can conclude that we are considering a set of Lagrangians with a parameter $\xi$. This point is important as it shows that in common tensor formalism only the sum $(\alpha+\beta)$ has a direct influence on dynamics. However, gauge invariance gives a restriction on $\xi$. Gauge phase multiplier of wave function can only be scalar because if it had a pseudoscalar part phase multipliers $\exp(\gamma_5 \kappa), \kappa\in \mathbb{R}$ would not cancel each other in  $\bar{\Psi}\partial \Psi $ because ${\exp(\gamma_5 \kappa)}^\dagger=\exp(\gamma_5 \kappa)$. Thus, we can only compensate the scalar gauge transform of $qA$, and if $qA$ has a pseudoscalar part, transformation of only scalar part would break constraints on $A$ and/or $q$. That inevitably gives $\alpha+\beta=2\pi n$ or $\xi=2 \pi n$ ($n \in \mathbb{Z}$). And we are back to the usual electrodynamics, as $\xi=2 \pi n$ gives us the Lagrangian (\ref{L0}). Physically it means that the gauge invariance forbids existence of observable magnetic charges. But there is a subtle point here: we can still use (\ref{L}) and its dynamic variables but with condition
\begin{eqnarray}
\xi=2 \pi n, \label{cond}
\end{eqnarray} 
that can take many forms in terms of expanded variables, e.g. $\frac{q^2(A\tilde{A})}{A^2 (q^*q)}=1$. Therefore there is a certain symmetry (not in the sense of Noether's symmetries): we are able to perform the following transforms
\begin{eqnarray}
q \raD e^{i\gamma_5\phi} q, \nonumber \\
A \raD e^{-i \gamma_5\phi} A \label{D}
\end{eqnarray}
without changing the Lagrangian (in fact, we are changing $\alpha \raD \alpha+\alpha' , \beta \raD \beta+\beta', \alpha'+\beta'=2 \pi n, \xi \raD \xi) $. These are duality transforms. But instead of transforming $F$ and $j$ as it is usually done in Maxwell's equations this way we can transform field itself and corresponding charge in Lagrangian. 
As we have proved, equation (\ref{L}) with the condition (\ref{cond}) is equivalent to (\ref{L01}) but has the symmetry (\ref{D}). Thus, the equation of motion derived from (\ref{L}) under condition  (\ref{cond}) and after returning to initial variables must give the same equation (\ref{max}), but it also must be dually symmetrical. Such an equation is
\begin{eqnarray}
\partial \mathcal{F}=j q. \label{eq}
\end{eqnarray}
 It can be split into vector and pseudovector parts adding and subtracting $\gamma_5(\ref{eq})\gamma_5$ and (\ref{eq}), from which it is obvious that pseudoscalar part of $q$ plays a role of magnetic charge, whereas the scalar one is the usual electric charge. Also, if we denote magnetic charge as $g$ then we get $\alpha=\arctan\frac{e}{g}$.
\par But despite all of that, the only new thing we achieved here is that the duality symmetry is made a symmetry of Lagrangian (we would like to note again that this is not a symmetry in terms of Noether's theorem because it changes parameters of the theory) and rewrote duality transform in terms of a field and corresponding charge.  But all of this is necessary to obtain local duality symmetry and understand the way it was obtained. 
\par Concluding this section let us clarify once more mathematical structure of our proposal. Initially, we had a theory in which there were variable $a$ and a parameter $e$. Then, we expanded the range of their values: $A=\exp{(i \gamma_5\beta)} a$, $q=e \cdot \exp{(i \gamma_5\alpha)}$. To make a good visual representation one may see new range of values as a plane $(\alpha, \beta)$, where set of points $(2\pi n, 2\pi n)$ is the initial theory. Later, we slightly modified the structure of obtained Lagrangian so that to be equivalent to the initial theory it is not  necessary to stay in $(2\pi n, 2\pi n)$, but rather stay on the set of lines $\alpha+\beta=2 \pi n$ parameterized through duality angle $\phi$, and moreover, gauge invariance implies the parameters to stay on these lines. The next step is to make these parameters local.
\end{section}

\begin{section}{Local duality symmetry}
\par Following the same logic as in gauge theories, we may allow $\alpha$ and $\beta$ (or, equivalently, $\phi$) be parameterized through spacetime coordinates and try to preserve the symmetry. That will make us, as usually, to change common derivatives of $A$ to covariant ones, $DA=(\partial + i\gamma_5 g_1 W)A$, here $g_1$ is coupling constant of $W$ and $A$. $W$ can be either a pseudovector field as in works of Tiwari \cite{Tiwari2011OnElectromagnetism}, \cite{Tiwari2011RoleInsulators},\cite{Tiwari2015AxionPerspective} or a pseudoscalar field derivative $\partial \phi$. We will focus on the second case as it is more complicated, and then briefly discuss the first case by analogy. The localized Lagrangian is then
\begin{eqnarray}
\mathcal{L}=\frac{1}{4} tr \left[ - jqA + \frac{1}{8}\frac{q^*}{q} ( (DA)^2 + (\widetilde{DA})^2 - \{DA,\widetilde{DA}\}_+ ) \right]. 
\end{eqnarray}
After getting back to old variables and taking into account $tr[\{a, b\}_+]=tr[2ab]$,
\begin{eqnarray}
\mathcal{L}=\frac{1}{4} tr \left[ - eja + \frac{1}{8}( (Da)^2 + (\widetilde{Da})^2 - 2 Da\widetilde{Da}) \right]. 
\end{eqnarray}
Introducing the notation $(\partial \phi)A -\widetilde{((\partial \phi)A)}=\Phi=\Phi_{\mu\nu} \gamma^\mu \gamma^\nu$, we get
\begin{eqnarray}
\mathcal{L}=\frac{1}{4} tr \left[ - eja + \frac{1}{8}( F^2 + g_1^2\Phi^2 + 2i\gamma_5 g_1\Phi F) \right]. 
\end{eqnarray}
Our $\Phi$ is not gauge invariant, and thus we could conclude that the local duality symmetry is not possible. But there is a possibility to compensate gauge transform $a\rightarrow a+\partial f\equiv a'$  with a duality transform $\partial \phi \rightarrow \partial\phi +\partial \chi$ in such a way that $\Phi$ remains invariant. We just need to demand 
\begin{eqnarray}
(\partial \phi)a=(\partial \phi + \partial \chi)(a+\partial f)
\end{eqnarray} 
and solve this as an equation for $\partial \chi$. The solution is  
\begin{eqnarray}
\partial_\mu \chi =-(\partial_\mu \phi) \frac{(\partial_\nu f) a'^\nu}{a'^2}. \label{comp}
\end{eqnarray}
This way, it is possible to preserve both gauge and duality symmetry. Alternatively, we could require $tr\left [ \Phi^2 \right ]=0$. That would lead us to 
\begin{eqnarray}
(\partial \phi)^2 a^2 = ((\partial_\mu \phi) a^\mu)^2,
\end{eqnarray} which represents a constraint of  $\partial_\mu \phi = y(x) a_\mu$, where $y(x)$ is an arbitrary function that changes under duality transform. Another, though weaker, approach, is to say that $\Phi^2$ is multiplied by $g_1^2$ and if $g_1$ is small enough we can neglect this term.
Now, let us take trace:
\begin{eqnarray}
\mathcal{L}=- e j^{\mu} a_{\mu}-\frac {1}{4} F_{\mu\nu}F^{\mu\nu}+\nonumber \\ +\frac{1}{2}g_1\epsilon_{\alpha \beta \mu \nu}(\partial^\alpha \phi) a^\beta F^{\mu\nu} + g_1^2(\partial \phi)^2 a^2 -g_1^2 ((\partial_\mu \phi) a^\mu)^2 . \label{full}
\end{eqnarray} 
The issue is that we cannot simply add the kinetic term $(\partial \phi)^2$ of $\phi$ (pseudoscalar field) to our Lagrangian, because it must obey the local duality transform $\partial \phi \rightarrow \partial\phi +\partial \chi$. To fix this one can use the same method that is used in some spontaneously broken symmetry models, namely, we add to Lagrangian (\ref{full}) additional pseudovector massive field $k$ transformed under the duality transform as $k \rightarrow k + \partial \chi/g_2$ and its interaction with $\partial \phi$. Our final Lagrangian takes thus the form
\begin{eqnarray}
\mathcal{L}=- e j^{\mu} a_{\mu}-\frac {1}{4} F_{\mu\nu}F^{\mu\nu}+ \nonumber \\+\frac{1}{2}g_1\epsilon_{\alpha \beta \mu \nu}(\partial^\alpha \phi) a^\beta F^{\mu\nu} +g_1^2 (\partial \phi)^2 a^2 - g_1^2((\partial_\mu \phi) a^\mu)^2 + \nonumber \\ + \frac{1}{2}(\partial \phi)^2+\frac{1}{2}g_2^2 K^2 - g_2\partial_\mu \phi K^\mu - \frac{1}{4}K_{\mu\nu}K^{\mu\nu}.
\end{eqnarray}
Here $K_{\mu\nu}=\partial_\mu k_\nu - \partial_\nu k_\mu$. The last four terms can be rewritten as 
\begin{eqnarray}
\mathcal{L}_k=\frac{1}{2}(\partial \phi)^2+\frac{1}{2}g_2^2 K^2 - g_2\partial_\mu \phi K^\mu - \frac{1}{4}K_{\mu\nu}K^{\mu\nu}=\nonumber \\ =\frac{1}{2}g_2^2(k_\mu - \frac{1}{g_2}\partial_\mu \phi)^2 -\frac{1}{4}K_{\mu\nu}K^{\mu\nu} . \label{Lk}
\end{eqnarray}
And it is obviously invariant under the duality transform. Hence, $\partial\phi$ plays a role of longitudinal component of $k_\mu$.
\par Alternatively, if we demand $W$ to be pseudovector field, we just need to replace $\partial\phi$ with $W$ everywhere up to (\ref{full}). The resulting Lagrangian is 
\begin{eqnarray}
\mathcal{L}=- e j^{\mu} a_{\mu}-\frac {1}{4} F_{\mu\nu}F^{\mu\nu}+\nonumber \\+\frac{1}{2}g_1\epsilon_{\alpha \beta \mu \nu}(W^\alpha) a^\beta F^{\mu\nu} + g_1^2(W)^2 a^2 -g_1^2 (W_\mu a^\mu)^2 . 
\end{eqnarray}
This case does not have the kinetic term issue that (\ref{full}) had. We can add $W_{\mu\nu}W^{\mu\nu}$, where $W_{\mu\nu}=\partial_\mu W_\nu - \partial_\nu W_\mu$, but then the field $W$ must be massless. Thus we can conclude, that if we choose $W$ to be pseudovector we obtain massless pseudovector field. If we choose it to be pseudoscalar derivative it becomes longitudal component of massive pseudovector field as it follows from (\ref{Lk}). Both options would require to compensate gauge transforms with local duality transforms (\ref{comp}).   
\end{section}

\begin{section}{Conclusion and perspectives}
	We constructed an extension of usual electrodynamics which includes the so-called "Local Duality Symmetry" (LDS). Our work was motivated by paper \cite{Katz1965} and the observation that globally symmetric electrodynamics in tensor variables has exactly the same Lagrangian as the usual one. A natural question emerges concerning observable consiquences of LDS. We can think of three possible options. The first one is the most simple - $g_1$ and $g_2$ are so small that we cannot observe them within current technological level. This would make $k$ field a good candidate for hidden mass, known as the dark matter, as it does not interact with anything except for $\partial\phi$, which in its turn interacts with a photon. This would cause observable effects of $k$ on photon dynamics of the order of $g_1g_2$. The second option is that in electroweak interaction this symmetry is somehow broken. But this is a matter of further research. And the third one might be related to some problems in quantum field theory  (e.g. non-renormalizability of the theory) that could occur during quantization. 
    \par Another interesting feature is that such procedure is feasible in quantum chromodynamics. Instead of electric charge one should take coupling constant $g_s$, and instead of 4-potential - the gluon field potentials $a_\mu^j$. Expanded variables for them we will denote as $q_s$ and $A^j$, they will transform in the same way as $q$ and $A$ ( all 8 gluon fields must be transformed simultaneously). Because the gluon field strength tensor contains coupling constant in it, it would transform exactly as electrodynamic field strength tensor:
    \begin{eqnarray}
    G^j=\partial A^j - \widetilde{\partial A^j} - q_s f^j_{ik}A^i A^k \raD e^{i\gamma_5 \phi}G^j.
    \end{eqnarray}The localization will also provide pseudoscalar field that can be associated with sort of a massless axion and a pseudovector field. But this is also a matter for further study. Considering the fact that it deals with axion-like field and so-called $\theta$-term, this path could lead to a new model for solution of strong CP-problem.
\end{section}
\bibliography{references}{}
\bibliographystyle{unsrt}
\end{document}